\documentclass[sigconf]{acmart}
\newif\ifprintcomments

\printcommentstrue

\usepackage{subcaption}

\usepackage{tikz}

\usepackage{amsthm}       % theorem/lemma/corollary environments + \qed + proof env

\usepackage{enumitem}   

\begin{document}

\title{ZODIAC: Zero-shot Offline Diffusion for Inferring Multi-xApps Conflicts in Open Radio Access Networks}

\author{Zeyu Fang}
\email{joey.fang@gwu.edu}
\affiliation{
  \institution{George Washington University}
  \city{Washington}
  \state{D.C.}
  \country{USA}
}

\author{Shu Hong}
\email{shu.hong@gwu.edu}
\affiliation{
  \institution{George Washington University}
  \city{Washington}
  \state{D.C.}
  \country{USA}
}
% \affiliation{%

\author{Huu Trung Thieu}
\email{huu_trung.thieu@nokia-bell-labs.com}
\affiliation{
  \institution{Nokia Bell Labs}
  \city{Murray Hill}
  \state{NJ}
  \country{USA}
}

\author{Nakjung Choi}
\email{nakjung.choi@nokia-bell-labs.com}
\affiliation{
  \institution{Nokia Bell Labs}
  \city{Murray Hill}
  \state{NJ}
  \country{USA}
}

\author{Tian Lan}
\email{tlan@gwu.edu}
\affiliation{
  \institution{George Washington University}
  \city{Washington}
  \state{D.C.}
  \country{USA}
}
\renewcommand{\shortauthors}{Fang et al.}

\begin{abstract}
Open Radio Access Network (O-RAN) enables network control through multi-vendor xApps operating both within and across layers, subnets, and domains, whose concurrent execution can trigger conflicts that are latent during the development phase. Existing conflict management approaches rely heavily on joint-execution data, which is often unavailable in practice. To address this limitation, we formalize a novel problem termed conflict reasoning, which involves identifying conflict-inducing conditions given only marginal datasets from each individual xApp. We propose ZODIAC, a three-stage framework for zero-shot conflict condition inference that comprises uncertainty-aware surrogate model training, trajectory-level diffusion training, and compositional guided denoising for efficient, physics-constrained, and reliable condition search. We derive a theoretical lower confidence bound showing that the compositional reasoning in ZODIAC serves as a principled surrogate for true conflict severity, with the epistemic penalty directly controlling the approximation gap. We evaluate ZODIAC on both the lightweight Mobile-Env platform across all three O-RAN Alliance conflict types (direct, indirect, and implicit) and a realistic NS-O-RAN-Flexric simulator. ZODIAC consistently outperforms baseline condition search methods, achieving over 20\% higher True Positive Rate at Top-20, substantially stronger Spearman rank correlation, greater scenario diversity, and competitive computational efficiency. Ablation studies confirm the necessity of each guidance component, with epistemic uncertainty penalties proving essential for filtering spurious conflicts. To the best of our knowledge, ZODIAC is the first framework in O-RAN that enables conflict reasoning from marginal offline data without requiring any joint-execution traces.
\end{abstract}





\maketitle

%% ----- 10 pages including references -----

\section{Introduction}

%\zeyu{I think we may call this problem another name instead of the term ``conflict detection''. In my understanding, conflict detection means to detect the conflict existence/types based on the (sequence of) system status, while our problem is to figure out the specific conditions(external inputs) that directly cause the conflict with the conflict definition given. For now I'll refer as to \problem{}} 
%\Tian{I think it can be called conflict reasoning, as it requires extracting the behavior logic and analyzing the models.}

%------- First Paragraph (background)

\begin{figure}[!t]
\centering
\includegraphics[width=\linewidth]{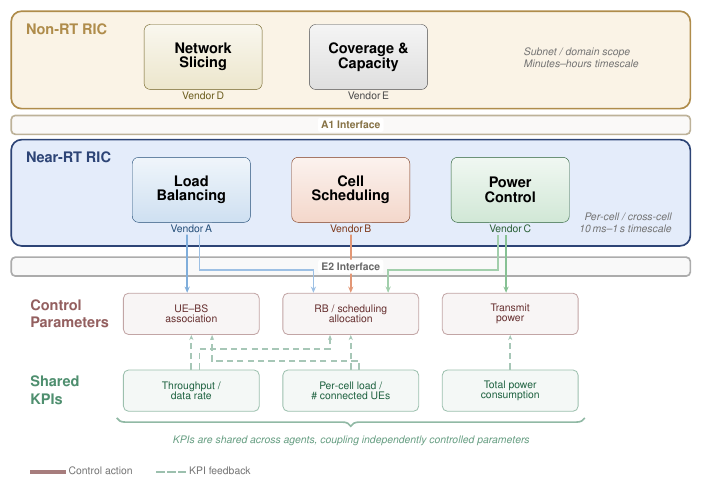}
\caption{O-RAN architecture with independently developed xApps across Non-RT and Near-RT RIC layers.
%Solid arrows denote control actions. Dashed arrows denote KPI feedback coupling independently controlled parameters. 
Three conflict types can arise that manifest only after joint deployment due to isolated development of xApps: 
(i) direct, when xApps issue divergent commands to the same parameter (e.g., Load Balancing and Cell Scheduling both writing to RB allocation) (ii) indirect, when disjoint parameters affect shared KPIs, so one xApp's action degrades another's objective (e.g., transmit power changes altering the throughput that Load Balancing observes) (iii) implicit, when each xApp individually satisfies a system constraint yet their combined actions violate it.}
\label{fig:oran-arch}
\vspace{-0.7cm}
\end{figure}

Open Radio Access Network (O-RAN) is reshaping next-generation wireless systems by disaggregating the RAN into open, interoperable components and exposing standardized interfaces for programmable control~\cite{polese2023understanding}.
At the center of this architecture, the Near-Real-Time RAN Intelligent Controller (Near-RT RIC) hosts third-party applications known as xApps, each optimizing specific network functions such as load balancing, power control, or quality-of-experience (QoE) management.
As illustrated in Figure~\ref{fig:oran-arch}, these xApps are typically developed in isolation by different vendors and deployed across overlapping or coupled network layers, functionalities, and domains.
Each xApp observes shared Key Performance Indicators (KPIs) through the E2 interface and issues control actions over parameters that may be shared with or physically coupled to those of other xApps.
Because xApps are typically designed, trained, and validated in isolation, their collective behavior under concurrent deployment is typically unknown prior to execution.
% \shu{Add a network architecture diagram at the beginning of the introduction section:
% }

% \shu{Source file for Fig.1 can be found in figures/oran-fig.tex}

% A critical challenge is thus introduced: xApps

This architecture introduces a structural challenge: xApps
%, often based on opaque machine-learning models, 
do not guarantee amicable collective behavior.
The concurrent execution of such xApps can trigger (latent) conflicts that manifest in multiple forms.
The O-RAN Alliance categorizes these into three types~\cite{oran_conflict_mitigation_2024}: (i)~\emph{direct conflicts}, where multiple xApps issue divergent commands over the same control parameter, (ii)~\emph{indirect conflicts}, where xApps controlling different but physically coupled parameters drive a shared KPI in opposite directions, causing unexpected performance degradation, and (iii)~\emph{implicit conflicts}, where each xApp individually satisfies local constraints, yet their combined actions violate a hidden system-level constraint that only surfaces under joint deployment.
These conflicts can compromise network stability, efficiency, and reliability~\cite{adamczyk2023challenges}, and their severity grows as future networks host an increasing number of autonomous, AI-driven xApps within or across layers, subnets, and domains.

%------- Second (motivation, existing works and their limitations)

Existing conflict management methods rely on extensive joint-execution data:
%follow a detection-then-mitigation paradigm: 
conflicts are first identified by analyzing xApp interactions from joint-execution data collected by running multiple xApps simultaneously in a digital twin or live network, and then resolved through prioritization, whitelisting, or policy coordination~\cite{del2025pacifista,giannopoulos2025comix,wadud2026ai}.
However, the assumption of extensive joint-execution data may not hold in practice.
%this paradigm relies on an implicit assumption that is often violated in practice: joint-execution data is readily available.
%In reality, 
xApps from different vendors are rarely co-tested prior to development; constructing a high-fidelity digital twin that captures the collective behavior of any combinatorial subsets of xApps is costly, and evaluating untested xApp combinations on live networks risks unpredictable service degradation.
Moreover, latent conflicts may manifest only under specific conditions (e.g., specific mobility patterns or traffic loads), making brute-force data collection inefficient.
The black-box nature of neural-network-based xApps further precludes deriving conflict-inducing conditions through purely logical or analytical reasoning.
To the best of our knowledge, no existing method addresses the problem of inferring conflict-inducing conditions from xApps' individual behavioral data without joint-execution traces.

To address this gap, we consider a novel problem termed conflict reasoning, formulated as zero-shot conflict condition inference.
We consider access to only datasets that are collected from individual deployment, capturing marginal trajectories under a single active xApp. %marginal offline datasets, each collected from an individual xApp deployment and recording network trajectories under a single active xApp. 
Given these marginal datasets and a known conflict criterion, the objective is to reason about the conflict-inducing conditions (\emph{i.e.}, the initial network state and the exogenous variable sequence, such as user mobility) that are most likely to trigger conflicts with joint deployment. %These conditions consist of the initial network state and a sequence of exogenous variables such as user mobility patterns or traffic arrivals.
The inferred conditions can then be replayed in a simulator or testbed to reproduce validation and worst-case analysis, enabling targeted diagnosis and downstream mitigation before real deployment. 
This problem is challenging for three reasons. First, the trajectories under concurrent xApp execution are never observed, as the available data contains only marginal behaviors from isolated execution/testing. Second, the xApps are treated as black-box policies, and their interactions cannot be derived from their opaque internal logic. Third, naively stitching together independently learned marginal models can produce physically implausible trajectories, leading to spurious conflict discoveries that are artifacts of model error rather than genuine cross-xApp interaction.

Our key insight is that, although joint xApp behavior is unobservable, the space of physically feasible network trajectories is highly structured (e.g., due to physical rules and interaction laws) and can be learned from marginal data. By modeling this structure as a trajectory distribution and combining it with independently learned policy likelihoods and conflict objectives, we can construct a principled surrogate for the unobserved joint-conflict distribution, enabling effective search for conflict-inducing scenarios without requiring joint execution data. Building on this insight, we propose \textbf{ZODIAC} (\textbf{Z}ero-shot \textbf{O}ffline \textbf{D}iffusion for \textbf{I}nferring multi-x\textbf{A}pps \textbf{C}onflicts), a three-stage framework for efficient conflict reasoning:
\begin{enumerate}

    \item \emph{Surrogate model training from marginal data.}  We train differentiable neural-network ensembles on the marginal data to simulate each xApp's policy and the network state-transition dynamics. Ensemble disagreement provides calibrated uncertainty estimates that quantify prediction reliability.

    \item \emph{Trajectory diffusion prior.} 
    Single-step surrogate models incur compound errors over long horizons, making autoregressive rollouts unreliable and prone to producing physically implausible trajectories.
    We build an unconditional diffusion model to learn a distributional prior that captures temporal coherence, valid feature correlations, and macro-level physical constraints of the network environment.

    \item \emph{Compositional guided search.} 
    %Because the xApps are black boxes, their joint interaction cannot be derived directly. Instead
    We steer the reverse diffusion process with scale-free compositional guidance that combines a conflict-maximizing objective with physical consistency and epistemic uncertainty penalties.
    This enables efficient search over the high-dimensional space of conditions while filtering out spurious conflicts due to model fault.
\end{enumerate}

A guaranty derived in Theorem~\ref{thm:lcb} shows that the compositional guidance is a principled surrogate for a lower confidence bound on the true conflict severity, with the epistemic penalty controlling the approximation gap.
Our main contributions are summarized as:
\begin{itemize}
    \item \textbf{Problem formalization.} We formalize conflict reasoning as a zero-shot condition inference problem, rigorously defining the assumptions on marginal data and the optimization objective. To our knowledge, this is the first formalization of the conflict reasoning problem, applicable to O-RAN and shared architectures with co-existing agents, i.e., xApps.

    \item \textbf{ZODIAC framework.} We propose ZODIAC, a three-stage framework that combines uncertainty-aware surrogate model training, trajectory-level diffusion priors, and compositional guidance to discover conflict-inducing conditions from only marginal datasets. A theoretical lower bound on confidence is derived to support the proposed framework.

    \item \textbf{Comprehensive evaluation.} 
Experiments on both Mobile-Env (covering all three conflict types) and the realistic NS-O-RAN-FlexRIC simulator show that ZODIAC outperforms Random Search, Backpropagation Through Time, and the Cross-Entropy Method baselines, achieving over 20\% higher True Positive Rate at Top-20, substantially stronger Spearman rank correlation ($\rho \approx 0.48$ vs.\ $\leq 0.34$), and greater scenario diversity.
    Ablation studies further verified the effectiveness of the compositional guidance.
    %We conduct extensive experiments and ablation studies on both Mobile-Env (covering all three conflict types) and the realistic NS-O-RAN-Flexric simulator. ZODIAC significantly outperforms baselines across multiple metrics, including accuracy, severity ranking, diversity, and computational efficiency.
\end{itemize}

\section{Related Works}

\textbf{Conflict Management in O-RAN.}
Existing works on O-RAN conflict management are dominated by the detection-mitigation framework~\citep{oran_conflict_mitigation_2024}, with approaches detecting and classifying conflicts by evaluating xApps in the digital twins and resolving them through prioritization or whitelisting~\citep{giannopoulos2025comix, del2025pacifista}. 
More recent methods improve this by modeling dependencies among xApps, parameters, and KPIs as graphs~\cite{bakri2025mitigating, zolghadr2025learning, wadud2025xapp}. The adoption of game theory~\citep{wadud2023conflict} and gradient optimization~\citep{cinemre2024gradient} has enhanced mitigation performance in a theoretical example scenario.
A second line of work focuses on mitigation through AI-based methods, including reinforcement learning~\citep{cinemre2025xapp, wadud2026ai}, policy distillation~\citep{erdol2025xapp}, and explainable AI~\citep{varshney2025explainable}. 
However, these methods ignore the inherent difficulties involved in obtaining joint datasets in the first place.
Another closely-related work~\citep{del2026predicting} adopts a distinct approach to model the policy of xApps from their separate offline profiles and predict the extent of conflict impacts given a specific scenario. Different from that, ZODIAC proposed in our work generates possible conflict scenarios that maximize the probability of conflicts, which can be used for worst-case analysis and targeted mitigation.

\noindent \textbf{Adversarial Scenario Generation and Falsification.}
A growing body of work in machine learning studies how to uncover rare but safety-critical failures. A representative line is AST~\citep{lee2020adaptive}, which formulates failure discovery as a Markov Decision Process (MDP) and solves it with reinforcement learning. More recent methods extend this idea with gradient-based or planning-based searches~\citep{elimelech2024falsification}. A follow-up work in autonomous driving trains adversarial agents specifically to induce failures~\citep{kulkarni2024crash}. A parallel work uses learned priors, including diffusion-based models, to synthesize realistic yet adversarial traffic scenes for evaluation~\citep{xie2024advdiffuser}.
Most of these share a common limitation: the search procedure relies on either offline data or a simulator repeatedly queried for interaction. But in the O-RAN, the joint execution data of multiple xApps can be unavailable, and an exhaustive simulator search is computationally prohibitive. Besides, the characteristics of network control, such as hybrid discrete-continuous state spaces, hard physical constraints, and multiple temporal scales, pose additional challenges.

\noindent \textbf{Diffusion Models for Trajectory Generation.}
Diffusion models generate the entire predicted trajectory through denoising, thus avoiding the compound errors induced by traditional models. A representative example is Diffuser~\citep{janner2022planning}, which shows that denoising can be interpreted as planning and that guidance can steer generated rollouts toward specific behavior patterns. Follow-up work~\citep{ajay2022conditional} extends this idea to show that conditional diffusion models can generate trajectories toward high-scoring regions. Parallel to this trend, compositional diffusion models provide a principled mechanism for satisfying multiple conditions simultaneously~\citep{du2023reduce}. Early compositional diffusion work showed that independently learned conditional scores can be used to generate combinations not seen during training~\citep{liu2022compositional}. More recent extensions~\citep{power2023sampling} transfer this idea to constrained planning. Systems such as TrajDiffuser~\citep{briden2025compositional} demonstrate that long-horizon trajectories can be generated concurrently while composing constraints with improved data efficiency. Inspired by these, we introduce a compositional diffusion model, which to the best of our knowledge marks the first such application in the field of network conflict management.

% I cannot find any publication information (even arxiv) of this paper
%- Open RAN Conflict Agents: Detecting and Mitigating xApp
% Conflicts with Generative Agents \url{https://xyzhang.ucsd.edu/papers/DC.Kwon_INFOCOM26_ORCA.pdf}

\begin{figure}
    \centering
    \includegraphics[width=\linewidth]{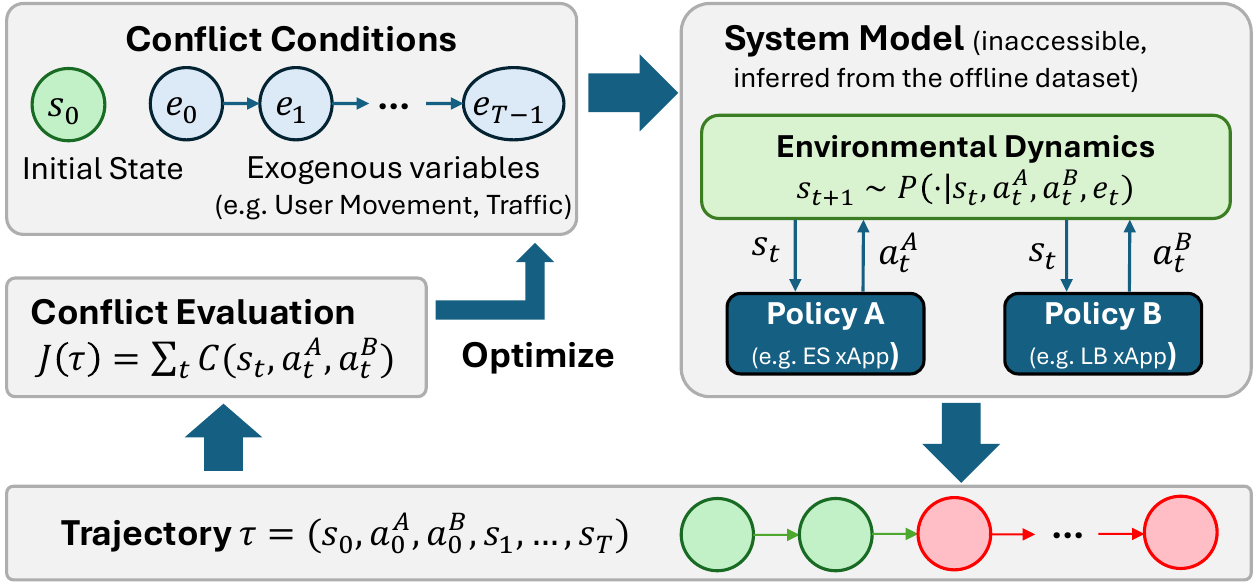}
    \caption{An illustration of the conflict reasoning problem.
Without true joint-execution data, we infer a surrogate system model entirely from offline marginal datasets. By simulating joint trajectories $\tau$ driven by input conflict conditions $(s_0, e_{0:T-1})$, the system iteratively optimizes the inputs to maximize the expected conflict severity $J(\tau)$.
    }
    \label{fig:problem_formulation}
    \vspace{-0.4cm}
\end{figure}

\section{System Model and Formulation}
\label{sec:problem}

% \shu{to-do: 1. conflict reasoning/discovery/inference?
% 2. trajectory feasibility prior? diffusion prior? would diffusion be  a bit narrow/low-level?
% }

% We will defined the problem of conflcit reasoning in this chapter.

% Include: the objectives and the constraints. The inputs and outputs. Definitiion of internal parameters/exogenous signals; Example.

In this section, we introduce the system model and formulate the conflict reasoning problem.
As illustrated in Figure~\ref{fig:problem_formulation}, we consider the general problem in which multiple policies operate concurrently within a system, optimizing different objectives. Such architectures arise naturally in modern network systems such as O-RAN.

\subsection{System Model}
Let $s_t \in \mathcal{S}$ denote the system state at time $t$, which refers to all the parameters and KPIs determining the current status of the network, such as base station load, user association, quality indicators, and resource allocation.
The states evolve over discrete time steps according to the inherent environmental dynamics of the system:
\begin{equation}
s_{t+1} \sim P(\cdot \mid s_t, a_t^A, a_t^B, e_t),
\end{equation}
where $a_t^A$ and $a_t^B$ are control actions produced by two independent control policies $\pi_A$ and $\pi_B$, and $e_t \in \mathcal{E}$ denotes exogenous variables.
These variables capture factors independent of the environmental states and dynamics while driving the evolution of the system, such as traffic arrivals and user mobility patterns.
We focus on two representative control policies, which iteratively sample an action given the current state at each time step, similarly to MDP:
\begin{equation}
a_t^A \sim \pi_A(\cdot \mid s_t), 
a_t^B \sim \pi_B(\cdot \mid s_t),
\end{equation}
These policies may correspond, for example, to energy-saving (ES) and load-balancing (LB) xApps in a cellular network. The two policies can act on both overlapping and non-overlapping configuration parameters, interacting through the shared network state.
\\
Given an initial state $s_0$ and a sequence of exogenous variables $e_{0:T-1}$, the joint execution of $\pi_A$ and $\pi_B$ within the system model induces a trajectory with a step length $T$:
\begin{equation}
\tau = (s_0,e_0,a_0^A,a_0^B,s_1,\ldots,s_T). \label{tau}
\end{equation}
% ====Potential structure: 1) introduce the systems, network dynamics (as above)

\subsection{Problem Formulation}

% 1) Background
Given the system model, our objective is to identify the conditions that trigger conflict outbreaks within a foreseeable time horizon $T$ when multiple policies operate concurrently.
Let condition $u = (s_0, e_{0:T-1})$ be defined as the combination of the initial state and the sequence of exogenous variables. 
Considering a scenario with two independent control policies, the joint execution under these conditions induces a trajectory distribution:
\begin{equation}
    \tau \sim P_\tau^{A,B}(\cdot \mid u) = P_\tau(\cdot \mid a_t^A \sim \pi_A, a_t^B \sim \pi_B, s_0, e_{0:T-1}).
\end{equation}
% 2) Define of conflicts, different conflict types.
To define the conflict outbreak, we assume that the extent of a conflict can be evaluated using a single-step conflict metric, denoted as $\phi_c(s_t, a_t^A, a_t^B) \ge 0$. 
Consequently, the cumulative conflict over a given trajectory is defined as
$J(\tau) = \sum_{t=0}^{T-1} \phi_c(s_t, a_t^A, a_t^B).$

Following the framework established by the O-RAN Alliance, the single-step conflict metric encompasses three distinct categories of policy interference. 
(i) Direct conflict occurs when multiple policies attempt to control the exact same parameters but generate divergent action instructions; (ii) Indirect conflict arises when policies control different parameters that are physically coupled through a shared KPI, which can be driven in opposite directions, thus resulting in unexpected values; (iii) Implicit conflict describes a more subtle phenomenon beyond these two. One typical scenario is where independent actions satisfy local requirements, but the joint execution of these actions violates a hidden capacity constraint of the system. Besides, the conflict metric $\phi_c$ should be carefully designed to isolate performance drops caused by specific inter-policy conflicts from those caused by inherent environmental adversity.

% 3) the objective
Ultimately, our objective is to discover the condition that maximizes the expected conflict severity:
\begin{equation}
    u^* = \arg \max_{u} \mathbb{E}_{\tau \sim P_\tau^{A,B}(\cdot \mid u)} \big[ J(\tau) \big].
\end{equation}

% 4) Constraints (the offline setting):
% \subsection{Offline Marginal-Deployment Setting}
However, solving this optimization problem presents a significant challenge when the joint-execution data is unavailable. 
% Controllers are typically developed and validated by independent vendors. Consequently, historical logs rarely contain trajectories where all advanced controllers are deployed simultaneously. Instead, operators typically possess datasets from marginal deployments, wherein one target controller operates actively while the system maintains the other parameters at a baseline configuration. 
Formally, for a given deployment environment, we assume the offline data only consists of two marginal sets: $\mathcal{D}_A = \{\tau \mid a_t^A \sim \pi_A(\cdot \mid s), a_t^B \sim \pi_B^0(\cdot \mid s)\}$ and $\mathcal{D}_B = \{\tau \mid a_t^A \sim \pi_A^0(\cdot \mid s), a_t^B \sim \pi_B(\cdot \mid s)\}.$ In these definitions, $\pi_A$ and $\pi_B$ represent the unknown target policies controlling actions $a_t^A$ and $a_t^B$, respectively, while $\pi_A^0$ and $\pi_B^0$ denote default policies executed through simple heuristics or fixed configurations. Because these sets exclusively capture the system dynamics under individual policy interventions, the joint trajectory distribution under concurrent execution remains strictly unobserved in the training data.
Therefore, the problem is not standard conflict detection from observed joint traces, but zero-shot conflict condition inference from marginal offline data.

\section{The Framework of ZODIAC}
\label{Sec:method}
In this section, we present ZODIAC, a three-stage framework that decomposes the conflict reasoning problem into trainable components and composes them at inference time. 
Stage~1 trains differentiable surrogate models of each xApp's policy and the shared network dynamics from the marginal datasets.
Stage~2 learns a trajectory-level diffusion prior, capturing temporal coherence and physical constraints.
Stage~3 steers the reverse diffusion process with compositional guidance that maximizes conflict severity, enforces dynamic consistency, and penalizes high-uncertainty regions.

%method in detail, which adopts a ``divide and conquer at training, compose at inference'' paradigm to address the core challenges via three stages. 
%\shu{May need to make the three components corresponding to each failure model.->
%Our method has three components, each addressing one failure mode of naive conflict search.
% Specifically:
% 1. Agent behavior + transition models -> infer local reactions from marginal data
% 2. Trajectory feasibility prior -> keep long-horizon scenarios realistic
% 3. Uncertainty-aware guided search -> find severe conflicts without exploiting model errors
% }

\subsection{Uncertainty-Aware Model Training from Marginal Offline Data}

In the first stage, we train uncertainty-aware models to approximate single-step policy behaviors and environmental transitions from the marginal datasets $\mathcal{D}_A$ and $\mathcal{D}_B$. Both are implemented as differentiable neural network ensembles, which provide gradients for guided searching and epistemic uncertainty estimates to prevent out-of-distribution (OOD) exploitation. 

For the policy models, we use behavioral cloning to approximate the marginal policies $\pi_A(a^A \mid s)$ and $\pi_B(a^B \mid s)$. Each policy is represented by an ensemble of $N$ independently initialized networks, whose prediction variance serves as an uncertainty estimate $U_\pi(s)$ that quantifies the model's confidence in a given state:
\begin{equation}
U_\pi(s) = \frac{1}{N} \sum_{n=1}^{N} \| \hat{a}^{(n)} - \bar{a} \|_2^2,
\label{U-a}
\end{equation}
where $\hat{a}^{(n)}$ is the action predicted by the $n$-th ensemble component and $\bar{a}$ is the mean.
For the dynamics, we train an ensemble of networks $\{T_{\phi_m}\}_{m=1}^M$ to predict the next state $s_{t+1}$ given the current state $s_t$, exogenous variable $e_t$, and joint actions $(a_t^A, a_t^B)$. The dynamics uncertainty is similarly estimated by the ensemble variance:
\begin{equation}
U_{\text{dyn}}(s_t, e_t, a^A, a^B) = \frac{1}{M} \sum_{m=1}^{M} \|\hat{s}_{t+1}^{(m)} - \bar{T}_\phi\|_2^2,
\label{U-d}
\end{equation}
where $\bar{T}_\phi$ is the mean prediction. We use cross-entropy loss for all action spaces, which are discrete. For the dynamics model, mean squared error (MSE) is adopted for continuous state features such as UE positions, while cross-entropy handles discrete features.

Together, these two models act as surrogate simulators during zero-shot inference, enabling single-step prediction without access to the true environment. 
%That said, cumulative prediction error over multiple steps can introduce substantial drift, limiting the reliability of long-horizon rollouts through these single-step models alone. This motivates the trajectory diffusion prior introduced in the next stage, which captures temporal coherence at the sequence level rather than relying on autoregressive model rollouts.
The uncertainty estimates $U_\pi$ and $U_{\text{dyn}}$ play a critical role here. Without them, gradient-based guidance during diffusion would steer the sampled conditions toward regions where the surrogate models are unreliable, causing model approximation errors to be falsely attributed to policy conflicts.

\subsection{Trajectory Diffusion Prior}

Single-step surrogate models trained in the last stage can suffer from compounding errors over long horizons, making autoregressive rollouts unreliable. To address this, we train an unconditional diffusion model over complete trajectories sampled from the marginal datasets. Rather than predicting one step at a time, this model learns a distributional prior over entire sequences, capturing temporal coherence, valid feature correlations, and the macro-level physical constraints of the environment. The resulting prior defines an in-distribution manifold that prevents the guided search in the next stage from producing physically implausible scenarios.

Let a full trajectory form be $\tau$ as defined in Eq. \eqref{tau}. We train an unconditional Denoising Diffusion Probabilistic Model (DDPM)~\cite{ho2020denoising}, parameterized as a noise predictor $\epsilon_\theta(\tau^k, k)$, on trajectories pooled from $\mathcal{D}_A \cup \mathcal{D}_B$. The forward process progressively corrupts a clean trajectory $\tau^0$ into isotropic Gaussian noise $\tau^K$ over $K$ diffusion steps, and the model is trained to reverse this corruption by minimizing the standard denoising objective:
\begin{equation}
\mathcal{L}_{\text{diffusion}} = \mathbb{E}_{\tau^0, k, \epsilon}\left[\|\epsilon - \epsilon_\theta(\tau^k, k)\|_2^2\right],
\end{equation}
where $k \sim \text{Uniform}(1, K)$ and $\epsilon \sim \mathcal{N}(0, I)$. During inference, the reverse process iteratively denoises a sample from $\tau^K \sim \mathcal{N}(0, I)$ back toward the learned data distribution.

A practical challenge in O-RAN is that state and action spaces often contain discrete features, while standard Gaussian diffusion operates in continuous space. We handle this through continuous relaxation: during preprocessing, discrete variables are mapped into $\mathbb{R}^d$ via learned embedding layers, and categorical actions are relaxed using the Gumbel-Softmax reparameterization. The diffusion model then operates entirely within this continuous latent representation. At the last few denoising steps, a hard projection operator $\Pi_{\mathcal{E}}$ maps the continuous outputs back to valid discrete parameters, enforcing feasibility constraints such as legal resource block assignments. This projection is also applied in inference steps.

% Because the diffusion prior is trained on real operational data, it implicitly encodes the physical regularities of the network environment. For instance, the exogenous condition sequences $e_{0:T}$ generated under this prior reflect realistic user mobility patterns rather than physically impossible transitions, such as instantaneous teleportation across cells. This grounding in observed data ensures that the conflict scenarios discovered through guided search in the subsequent stage remain practically meaningful.

\subsection{Compositional Energy Guided Search}

A generative model trained on $\mathcal{D}_A \cup \mathcal{D}_B$ merely simulates a mixture of marginal trajectory distributions with normal denoising steps and cannot represent the coupled dynamics that emerge under joint execution or be used to discover plausible trajectories with high conflict probabilities. To overcome this, ZODIAC steers the reverse diffusion process at inference time to sample from the unobserved joint conflict distribution $p(\tau \mid \pi_A, \pi_B, \phi_c)$. By Bayes' theorem, the score of this target distribution decomposes into the unconditional diffusion prior and a composite energy function $\mathcal{J}_{\text{guide}}(\tau)$ that encodes policy likelihoods and conflict objectives.

Since the intermediate diffusion iterates $\tau^k$ are corrupted by Gaussian noise, evaluating the surrogate models directly on $\tau^k$ produces invalid and numerically unstable gradients. We address this with Tweedie's formula~\cite{efron2011tweedie}, which provides a closed-form estimate of the clean trajectory $\hat{\tau}_0$ from any noisy iterate at step $k$:
\begin{equation}
\hat{\tau}_0 = \frac{\tau^k - \sqrt{1-\bar{\alpha}_k}\,\epsilon_\theta(\tau^k, k)}{\sqrt{\bar{\alpha}_k}}.
\end{equation}
All subsequent gradient computations are performed on this denoised estimate. We define three energy functions over $\hat{\tau}_0$, each serving a distinct role in the guided search.

The target energy $\mathcal{E}_{\text{target}}$ drives the trajectory toward high-conflict regions while ensuring that the actions remain consistent with the learned marginal policies:
\begin{equation}
\mathcal{E}_{\text{target}} = \sum_{t=0}^{T-1} \big( \log \pi_A(\hat{a}_t^A \mid \hat{s}_t) + \log \pi_B(\hat{a}_t^B \mid \hat{s}_t) + \phi_c(\hat{s}_t, \hat{a}_t^A, \hat{a}_t^B) \big),
\end{equation}
where $\phi_c$ is instantiated according to the target conflict. For direct conflicts, this corresponds to the KL divergence between policy action distributions; for indirect or implicit conflicts, it takes the form of continuous penalty functions measuring performance degradation or constraint violations directly caused by the joint policy.

The physics energy $\mathcal{E}_{\text{physics}}$ enforces dynamic consistency by penalizing deviations between the generated state sequence and the predictions of the surrogate transition model:
\begin{equation}
\mathcal{E}_{\text{physics}} = -\sum_{t=0}^{T-1} \|\hat{s}_{t+1} - \bar{T}_\phi(\hat{s}_t, \hat{e}_t, \hat{a}_t^A, \hat{a}_t^B)\|_2^2.
\label{e-p}
\end{equation}

The epistemic energy $\mathcal{E}_{\text{epistemic}}$ discourages the search from exploiting regions where the surrogate models are unreliable, penalizing trajectories that pass through high-uncertainty states:
\begin{equation}
\mathcal{E}_{\text{epistemic}} = -\sum_{t=0}^{T-1} \big(U_{\text{dyn}}(\hat{s}_t, \hat{e}_t, \hat{a}_t^A, \hat{a}_t^B) + U_\pi(\hat{s}_t)\big).
\label{e-e}
\end{equation}

A common failure mode in multi-objective guided diffusion is gradient domination, where one energy term overwhelms the others due to differences in numerical scale, causing the trajectory to collapse into physically impossible states. To prevent this, we normalize the gradient of each energy component to unit $L_2$ norm, yielding scale-free direction vectors $g_{\text{target}}$, $g_{\text{physics}}$, and $g_{\text{epistemic}}$. Specifically, let $\tilde{g}_i = \nabla_{\hat{\tau}_0} \mathcal{E}_i$ denote the raw gradient of each energy component with respect to the denoised trajectory estimate. The scale-free direction vectors are obtained by $L_2$ normalization:
\begin{equation}
g_i = \frac{\tilde{g}_i}{\|\tilde{g}_i\|_2 + \delta}, \quad i \in \{\text{target},\, \text{physics},\, \text{epistemic}\},
\end{equation}
where $\delta$ is a small constant for numerical stability.
The composite guidance direction is then:
\begin{equation}
d_{\text{guide}} = \gamma \cdot g_{\text{target}} + (1 - \gamma) \cdot \frac{g_{\text{physics}} + g_{\text{epistemic}}}{2},
\end{equation}
where $\gamma \in [0,1]$ governs the trade-off between conflict intensity and physical plausibility. The guided denoising step takes the form:
\begin{equation}
\tau^{k-1} = \text{DDPM\_Step}\big(\tau^k,\; \epsilon_\theta - \eta \sqrt{1 - \bar{\alpha}_k} \cdot d_{\text{guide}}\big).
\end{equation}

%After each reverse step, the hard projection operator $\Pi_{\mathcal{E}}$ is applied via $\tau^{k-1} \leftarrow \Pi_{\mathcal{E}}(\tau^{k-1})$ to enforce physical boundary constraints and snap continuous relaxations back to valid parameter domains. This ensures that the final generated scenario $e_{0:T}^*$ is directly executable in the target simulator.

\section{Theoretical Analysis}
% \begin{equation}
% s_{t+1} = f_0(s_t,e_t) + f_A(s_t,e_t,a^A_t) + f_B(s_t,e_t,a^B_t) + f_{AB}(s_t,e_t,a^A_t,a^B_t) + \xi_t
% \end{equation}
% If we can give assumption on the first two models (also uncertainties), then the found conflict should be bound?

% \shu{
% what about: 
% 1. first give a lemma for: joint controller interactions are not identifiable from marginal deployments alone (the true joint system cannot be uniquely determined) -> highlight the challenge, necessiate the uncertainty penalties for our method.\\
% 2.
% {\textbf{trajectory mismatch bound under the assumption bounded interaction}} and/or perhaps, robustness guarantee: Lower Confidence Bound
% }

\label{sec:theory}

We analyze ZODIAC along two axes: (i) we expose the structural
obstruction that makes joint conflict reasoning from marginal data
fundamentally hard, and (ii) we show that the compositional guidance
of Section~\ref{Sec:method} is a tractable surrogate for a Lower Confidence Bound
(LCB) on the true conflict severity. Throughout this section, we work
under a set of mild regularity conditions (A1)--(A4) that capture
\emph{bounded cross-policy coupling}, \emph{Lipschitz dynamics,
policies and conflict metric}, and \emph{calibrated ensemble
uncertainty}. Their precise statements, together with all proofs,
are deferred to Appendix~\ref{app:proofs}.

To make the source of difficulty explicit, we adopt the canonical
decomposition of the joint dynamics in
Eq.~(1):
\begin{align}
\label{eq:decomp}
s_{t+1} =\; & f_0(s_t,e_t) + f_A(s_t,e_t,a^A_t) + f_B(s_t,e_t,a^B_t) \notag\\
            & + f_{AB}(s_t,e_t,a^A_t,a^B_t) + \xi_t,
\end{align}
where $f_0$ encodes the baseline-only response, $f_A$ and $f_B$
isolate the marginal effects of each policy relative to their corresponding default opposite policy $\pi_B^0$ or $\pi_A^0$, and $f_{AB}$ is the residual coupling that vanishes
whenever \emph{either} action equals the default action. Under this
decomposition, the marginal datasets $\mathcal{D}_A,\mathcal{D}_B$
of Section~3.2 reveal $f_0,f_A,f_B$ but provide \emph{no} signal
about $f_{AB}$.

\smallskip
\noindent\textbf{Non-identifiability of the coupling term.}
The first result formalizes the data-collection bottleneck.

\begin{lemma}[Non-identifiability of $f_{AB}$]
\label{lem:nonid}
Under the boundary condition on $f_{AB}$, the trajectory
distributions induced on $\mathcal{D}_A\cup\mathcal{D}_B$ by any two
choices of the coupling term $f_{AB},\tilde f_{AB}$ are identical.
Consequently, $f_{AB}$ cannot be uniquely recovered from marginal
data regardless of dataset size.
\end{lemma}

Lemma~\ref{lem:nonid} has two implications for ZODIAC's design.
First, the bounded-coupling assumption (A1) is \emph{necessary}:
without a priori bound on $f_{AB}$, no finite-sample
guarantee on the joint conflict can exist. Second, any point-estimate
surrogate trained on marginal data implicitly sets $f_{AB}\!\equiv\!0$,
which is one feasible explanation among infinitely many; the
epistemic energy in Eq.~(14) is precisely the term that penalizes
regions where this aliasing dominates.

Let $\hat F:=\bar T_\phi$ denote the surrogate joint dynamics
ensemble, and let
$\hat\tau$ and $\tau$ denote the surrogate rollout (under $\hat F$
with the policy ensembles $\hat\pi_A,\hat\pi_B$) and the true joint
trajectory under condition $u=(s_0,e_{0:T-1})$, respectively. Define
the per-step uncertainty radius
\begin{equation}
\label{eq:r}
r_k(u) := \beta_T\sqrt{U_{dyn}^{(k)}} + 2L_a\beta_\pi\sqrt{U_\pi^{(k)}}
         + L_{AB} + \sigma_\xi,
\end{equation}
and the closed-loop contraction constant $L':=L_F+2L_aL_\pi$, where
the Lipschitz constants $L_F,L_a,L_\pi$, the coupling bound $L_{AB}$,
the calibration constants $\beta_T,\beta_\pi$, and the noise scale
$\sigma_\xi$ are all introduced in Appendix~\ref{app:proofs}. $U_{dyn}$ and $U_\pi$ are the epistemic uncertainties of the dynamics and policy models, which are estimated by the ensembles using Equation~\eqref{U-a} and \eqref{U-d}.

\begin{theorem}[Trajectory mismatch]
\label{thm:traj}
Under (A1)--(A4), with probability at least $1-2T\delta$,
\begin{equation}
\label{eq:traj-bound}
\mathbb{E}\|\hat s_t - s_t\|_2
\;\le\; \sum_{k=0}^{t-1}(L')^{\,t-1-k}\, r_k(u),\quad \forall t\le T.
\end{equation}
\end{theorem}

The bound separates the rollout error into (a) the model epistemic
terms $\beta_T\sqrt{U_{dyn}}$ and $\beta_\pi\sqrt{U_\pi}$, which
ZODIAC's epistemic penalty defined in Eq.\eqref{e-e} directly suppresses; (b) the
irreducible coupling bias $L_{AB}$ that no amount of marginal data
can shrink (Lemma~\ref{lem:nonid}); and (c) the aleatoric floor
$\sigma_\xi$. The geometric factor $(L')^{t-1-k}$ is the standard
worst case for step-by-step autoregressive surrogates; ZODIAC blunts it because
the diffusion generates entire trajectories simultaneously, and the physics
energy defined in Eq.\eqref{e-p} keeps generated states on the surrogate's reliable
manifold where (A4) applies. Lifting the per-state bound to the
scalar conflict via Lipschitzness of $\phi_c$ yields our main
guarantee.

Let $J^*(u):=\mathbb{E}_{\tau\sim P^{A,B}_\tau(\cdot|u)}[J(\tau)]$
denote the true expected cumulative conflict and
$\hat J(u):=\sum_{t=0}^{T-1}\phi_c(\hat s_t,\hat a^A_t,\hat a^B_t)$
its surrogate.

\begin{theorem}[Conflict Lower Confidence Bound]
\label{thm:lcb}
Under (A1)--(A4), define
\[
R(u) := L_c\!\sum_{t=0}^{T-1}\!\Bigl[
2\beta_\pi\!\sqrt{U_\pi^{(t)}} +
(1{+}2L_\pi)\!\!\sum_{k=0}^{t-1}\!(L')^{t-1-k} r_k(u)
\Bigr].
\]
Then with probability at least $1-2T\delta$,
\begin{equation}
\label{eq:lcb}
J^*(u)\ \ge\ \hat J(u)\,-\,R(u).
\end{equation}
\end{theorem}

\begin{corollary}[Justification of compositional guidance]
\label{cor:guidance}
Up to scale-free normalization (Eq.~15), ZODIAC's composite guidance
is a first-order surrogate for the LCB $\hat J(u)-\lambda R(u)$:
$\mathcal{E}_{\mathrm{target}}$ ascends $\hat J(u)$,
$\mathcal{E}_{\mathrm{epistemic}}$ descends the
ensemble-disagreement portion of $R(u)$, and
$\mathcal{E}_{\mathrm{physics}}$ together with the diffusion prior
keeps the search inside the manifold where (A4) is informative.
Hence any condition $u^\star$ returned by ZODIAC satisfies, with
probability at least $1-2T\delta$,
\begin{equation}
\label{eq:certificate}
J^*(u^\star)\ \ge\ \hat J(u^\star) - R(u^\star),
\end{equation}
i.e., the discovered conflict is non-trivial whenever
$\hat J(u^\star)>R(u^\star)$.
\end{corollary}

Eq.~\eqref{eq:certificate} also predicts the ablation results observed in the experiments: removing $U_{dyn}$ or $U_\pi$ inflates
$R(u)$ relative to $\hat J(u)$, so the surrogate ranking ceases to
lower-bound the true conflict, thus causing the True Positive Rate
to collapse.

\section{Experiments}
\label{sec:experiments}

To demonstrate the effectiveness of ZODIAC, we conducted extensive experiments across multiple scenarios in both lightweight and realistic simulated environments. We first detail the configurations, including environment settings, conflict scenario design, evaluation metrics, and baselines. We then present the main experimental results and ablation studies to validate our framework design.

\subsection{Environment Preparation}

\begin{figure*}
    \centering
    \includegraphics[width=0.85\linewidth]{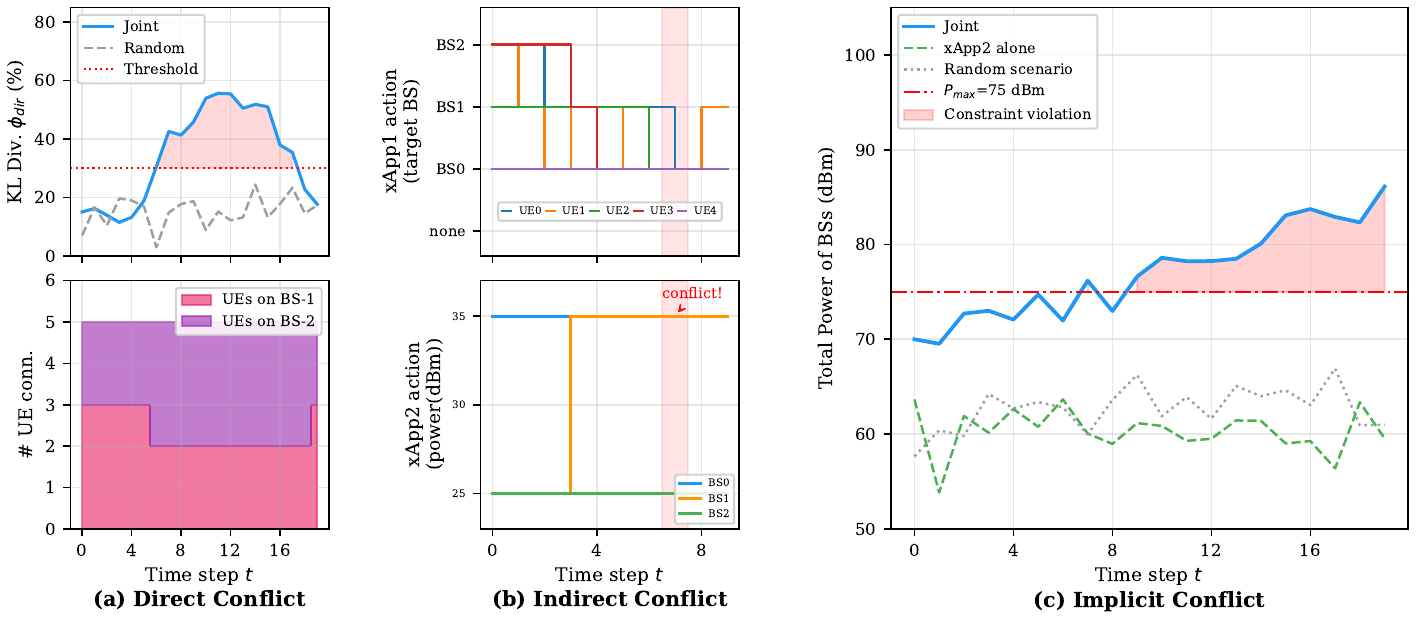}
    \caption{Visualization of the conflict cases for three different types in Mobile-Env environment. (a) The direct conflict occurs when the LB xApp and QoE xApp have distinct commands within the same time step, estimated by the KL Divergence between policy models. (b) The indirect conflict is defined as high-power BS with no connected UE, which can happen due to suboptimal cooperation between the UE-BS association xApp and Power control xApp. (c) The implicit conflict is defined as a violation of constraint (limited total power) due to the two xApps, which will not happen when running either xApp alone.}
    \label{fig:case}
    \vspace{-0.1cm}
\end{figure*}

Our experiments are conducted on two platforms of different fidelity levels. The first is Mobile-Env~\cite{schneider2022mobile}, a lightweight, open-source environment for wireless mobile networks. Mobile-Env models User Equipment (UE) moving in a two-dimensional area and connecting to Base Stations (BS). With a similar scenario set, the second environment is built on NS-O-RAN-Flexric~\cite{ns-o-ran-flexric}, a high-fidelity 5G O-RAN simulator that enables the deployment of xApps. Both platforms are adjusted to support the joint execution of multiple xApps and to record the necessary interfaces required by our framework.

\subsubsection{Mobile-Env Scenario Design}

Within the Mobile-Env, we construct three distinct scenarios, each corresponding to one of the conflict types. All three scenarios share a common network topology consisting of 3 BSs and 5 mobile UEs. Each UE is limited to connect to at most one BS at any time step. The default OkumuraHata channel model and log-based Shannon utility functions are adopted. The scenarios differ in policy objectives, action spaces, and conflict definitions, as illustrated in Figure~\ref{fig:case} and described below.

\textbf{Direct Conflict Scenario.}
Both xApps control the same parameter set as the UE-BS association, \emph{i.e.}, they share the same action space $\mathcal{A}_A = \mathcal{A}_B$ but optimize distinct objectives/KPIs:
\begin{equation}
    a^A, a^B \in \{\text{disconnect},\;\text{BS}_0,\;\text{BS}_1,\;\text{BS}_2\}
    \;\;\forall u,
\end{equation}
where $u$ is any one of the 5 UEs. Specifically, $\pi_A$ is the policy of a Load Balancing (LB) xApp that minimizes the maximum per-BS UE count and therefore tends to route UEs to lightly-loaded BSs. $\pi_B$ (QoE Maximization xApp) maximizes the aggregate data rate by concentrating UEs on the BS with the highest SINR, which is typically the nearest one. Because both xApps issue connection commands to the same UEs, their instructions are likely to be contradictory. The training procedures for these two policies are conducted independently in separate environments, using the classic RL algorithm PPO with different reward functions. The same procedures are applied for training the xApps in other scenarios. These initial policies are only used for data generation and real conflict verification, which are inaccessible to the ZODIAC framework. To clarify, the simulated policy ensembles in ZODIAC are another model trained to infer these true policies from their offline datasets. 

% \textit{Training environments and rewards.} The two policies are trained independently in separate environments:
% \begin{align}
%     \pi_A &: \texttt{mobile-conflict-Direct-xAppA-v0}, \nonumber\\
%       &\quad r_A = \bar{u}(s,a_A)
%         - 0.3 \cdot \frac{\max_b |\mathcal{C}_b|}{N_{\mathrm{UE}}}, \\[4pt]
%     \pi_B &: \texttt{mobile-conflict-Direct-xAppB-v0}, \nonumber\\
%       &\quad r_B = \frac{1}{|\mathcal{U}_{\mathrm{active}}|}
%         \sum_{u \in \mathcal{U}_{\mathrm{active}}} R_u(s, a_B),
% \end{align}
% where $|\mathcal{C}_b|$ denotes the number of UEs connected to BS~$b$, $R_u$ is the raw macro data rate (bps) of UE~$u$, and $\bar{u}$ is the mean Shannon-log-based utility scaled to $[-1,1]$. The concave utility in $r_A$ penalizes load concentration, while the linear reward $r_B$ explicitly incentivizes SINR-based concentration, thereby creating a fundamental tension between the two policies.

The conflict $\phi_{dir}$ in this scenario is defined as the inconsistency of actions from two policies within the same time step. The system process will continue with a priority for LB xApp over QoE xApp. To provide continuous conflict metrics estimates $\hat{\phi}_{dir}$ for guidance, given the trained policy ensembles $\{\hat{\pi}_A(\cdot\mid s),\,\hat{\pi}_B(\cdot\mid s)\}$, we define:
\begin{equation}
    \hat{\phi}_{\mathrm{dir}}(s_t,\hat{\pi}_A, \hat{\pi}_B)
    = 
      \frac{1}{N_{\mathrm{UE}}} \sum_{u=1}^{N_{\mathrm{UE}}}
      D_{\mathrm{KL}}\!\bigl(
        \hat{\pi}_A(\cdot \mid s_t,u)
        \;\|\;
        \hat{\pi}_B(\cdot \mid s_t,u)
      \bigr),
\end{equation}
In which $N_{UE}$ is the number of UEs. Therefore, a high KL divergence indicates that the two xApps are more likely to give conflicting commands for the UE-BS assignment under the same network state. % Importantly, this metric possesses a natural immunity to trivially adverse states: when the environment is so degraded that both policies converge to the same action (e.g., disconnecting all UEs), the KL divergence approaches zero, and such states are not falsely flagged as conflicts.

\textbf{Indirect Conflict Scenario.}
The two xApps control disjoint parameter sets, \emph{i.e.}, $\mathcal{A}_A \cap \mathcal{A}_B = \emptyset$. However, their joint actions can impact the shared dynamics with a combined effect, leading to performance degradation. More specifically, $\pi_A$ controls per-BS transmit power, while $\pi_B$ controls UE-BS association:
\begin{align}
      & a_t^{A,(b)} \in \{20\,\text{(low)},\; 30\,\text{(med)},\; 40\,\text{(high)}\} (\text{dBm}),
      \;\;\forall b\\
      & a_t^{B,(u)} \in \{\text{disconnect},\;\text{BS}_0,\;\text{BS}_1,\;\text{BS}_2\},
      \;\;\forall u.
\end{align}

Typically $\pi_A$ reduces the power levels of BSs with few connections to conserve energy. Under the OkumuraHata channel model, this could cause the received SNR at cell-edge UEs to drop below the service threshold. Meanwhile, $\pi_B$ maximizes the QoS while minimizing handover overhead. Therefore, in some cases, the joint effect could trap UEs on low-power BSs, severely degrading their utility, which is a conflict that is hard to infer from marginal observations.

% \textit{Training environments and rewards.}
% \begin{align}
%     \pi_A &: \texttt{mobile-conflict-Scenerio2-xApp2-v0},
%       \quad r_A = \bar{u} - \lambda_p \cdot E_{\mathrm{consumption}}, \\
%     \pi_B &: \texttt{mobile-conflict-Scenerio1-xApp1-v0},
%       \quad r_B = \bar{u} - \lambda_c \cdot N_{\mathrm{handover}},
% \end{align}
% where $\bar{u}$ is the mean scaled utility in $[-1,1]$, $E_{\mathrm{consumption}}$ quantifies the total energy expenditure, $N_{\mathrm{handover}}$ counts the number of handover events, and $\lambda_p = \lambda_c = 0.1$.

The per-BS indirect conflict definition $\phi^b_{\mathrm{ind}}$ is based on a directly observable pattern: a BS operating at minimum power while still serving a disproportionate number of UEs:
\begin{equation}
    \phi^b_{\mathrm{ind}}(s_t,a^A_t, a^B_t)
    = 
      \mathbb{I}\!\left[a^{A,(b)}_t= low\right]
      \cdot
      \mathbb{I}\!\left[\textstyle\sum_u \mathbb{I}[a_t^{B,(u)} = b] \geq \theta_c\right],
\end{equation}
with threshold $\theta_c = 2$. The overall conflict metric is defined as $\phi_{ind}(s_t,a^A_t, a^B_t) = \sum_b\phi^b_{ind}(s_t,a^A_t, a^B_t)$. The physical interpretation of this definition is clear: BS~$b$ operates at its minimum power level while $\theta_c$ or more UEs remain associated with it. This pattern emerges precisely when one xApp lowers power while the other simultaneously refuses to reroute the affected UEs.

\textbf{Implicit Conflict Scenario.}
The action spaces of two xApps are the same as in the indirect scenario, but the conflict is hidden: it only manifests under joint deployment because each policy individually obeys a constraint, yet their combined actions could exceed it. 
In isolation, each xApp is trained to maximize the QoE while satisfying the total transmit-power budget $P_{\max}$:
\begin{equation}
    \sum_{b=1}^{3} p_b^{(k)}(t) \leq P_{\max}, \quad k \in \{A, B\},
\end{equation}
where $p_b \in \{20, 30, 40\}\,\text{dBm}$ is mapped to $\{0, 1, 2\}\,\text{W}$ for summation. Under joint deployment, $\pi_A$ can steer more UEs toward a particular BS, while Policy~$B$ responds by boosting the power of that BS to satisfy the QoE. Their combined effect can cause unpredictable results, leading to exceeding $P_{\max}$.
Therefore, we define the conflict metric based on the constraint violation:
\begin{equation}
    \phi_{\mathrm{imp}}(s_t,a^A_t,a^B_t)
    = \max\!\Bigl(0,\; \sum_{b=1}^{3} p_b(t) - P_{\max}\Bigr).
\end{equation}
This function is strictly zero when the power budget is respected and increases proportionally with the severity of the violation. Since each individual policy satisfies the budget during marginal training, a nonzero $\phi_{\mathrm{imp}}$ unambiguously signals a conflict that arises exclusively from joint deployment.

% \textit{Training environments and rewards.}
% \begin{align}
%     \pi_A &: \texttt{mobile-conflict-Implicit-xAppA-v0},
%       \quad r_A = \bar{u} - \lambda_{\mathrm{ho}} \cdot N_{\mathrm{handover}}, \\
%     \pi_B &: \texttt{mobile-conflict-Implicit-xAppB-v0},
%       \quad r_B = \bar{u} - \lambda_p \cdot E_{\mathrm{consumption}}.
% \end{align}

\subsubsection{NS-O-RAN-FlexRIC Scenario}

To validate ZODIAC in a high-fidelity O-RAN setting, we establish an environment on NS-O-RAN-FlexRIC, which integrates the FlexRIC Near-RT RIC with the ns-3 mmWave module and the 5G-LENA NR module, providing a Non-Standalone 5G network with standardized E2 interfaces. The simulation features one LTE eNB and 5 mmWave gNBs, with 10 UEs moving in the area and being served through dual connectivity.

We deploy two xApps analogous to the direct conflict scenario: an LB xApp that distributes UEs across gNBs to equalize cell utilization and an Energy Saving xApp that reduces cell power or deactivates lightly-loaded cells to minimize energy consumption. Their concurrent execution on the same set of handover control parameters creates a direct conflict, which manifests as oscillatory handover decisions and degraded user throughput. The KPIs are collected through the standardized KPM v3 indication messages exchanged over the E2 interface, and the control actions are issued via RC v1.03 control requests. A major difference in NS-O-RAN-FlexRIC compared with Mobile-Env is that the xApps now have delayed decision-making and action execution: their actions are made based on a moving window average and network status over consecutive time intervals, while the transfer action is done by sending the signals first and is carried out by other modules. To deal with this, we integrate 10 time steps into one and merge all the corresponding UE movements and actions taken by the two xApps.

\subsubsection{Data Collection}
\label{sec:data_collection}

For each scenario, the data collection process follows the marginal deployment protocol established in Section~\ref{sec:problem}. After training both xApps, offline datasets $\mathcal{D}_A$ and $\mathcal{D}_B$ are collected by rolling out each policy for a fixed number of episodes. Specifically, for indirect and implicit conflict scenarios, since the actions of the two xApps do not overlap, a marginal policy is required to cover the action that would have been taken by the other xApp. In this setting, for the data collection process concerning the "UE-BS Association Control xApp," the marginal policy is configured to maintain a fixed power level; conversely, for the "Power Control xApp," the marginal policy employs a "most recently assigned" rule. 

During each episode, the initial UE positions and moving directions are randomly sampled, and UE speed is held constant. The environment is reset with a new random configuration at the start of each episode. Each trajectory record contains the exogenous variables $e_t$ (UE moving directions), the system state $s_t$, and the actions of both the active policy and the marginal policy. Crucially, no joint deployment data is collected at any point since the two xApps have never co-existed during data collection.

\subsection{Evaluation Metrics and Baselines}
\label{sec:metrics}

Our evaluation verifies true conflicts by applying the conditions to the ground-truth simulator. We adopt the following metrics:

\textbf{True Positive Rate at Top-K (TPR@K).} The generated scenarios $\hat\tau$ are ranked by their estimated conflict energy $J(\hat\tau)$ in descending order. The conflict conditions of the top-$K$ scenarios are then extracted and verified. A condition is counted as a true positive if its conflict metric $\phi_c$ exceeds a predefined threshold. TPR@$K$ reports the fraction of true positives among them.

\textbf{Spearman~$\rho$.} Beyond TPR@K, we measure whether the energy of the simulated trajectory correctly ranks scenarios by their true conflict severity. Spearman's rank correlation coefficient is computed between the offline energy $J(\hat\tau)$ and the corresponding simulator-verified conflict metric $J(\tau | s_0 = \hat s_0, e_{0:T-1} = \hat e_{0:T-1})$. A high $\rho$ indicates that the method not only finds conflicts but also correctly reasons about them and predicts the scenario.

\textbf{Computational Efficiency.} We report the wall-clock time required by each method to generate its full set of candidate scenarios. %This metric is critical for practical deployment, where conflict assessment must be performed within reasonable time budgets.

\textbf{Diversity.} To ensure that the method discovers a broad range of conditions rather than being confined to a few failure modes, we report the distribution of conflict severity across all generated scenarios and examine the diversity of discovered conflict patterns.

We compare ZODIAC against three baselines that represent different paradigms for failure mode searching:

\textbf{Random Search (RS).} Exogenous condition sequences $e_{0:T-1}$ and initial states $s_0$ are sampled uniformly at random. This baseline establishes the base rate of conflict occurrence under natural conditions and serves as a lower bound on search effectiveness.

\textbf{Backpropagation Through Time (BPTT).} This baseline directly backpropagates gradients through the surrogate model across all time steps to optimize the conditions. Starting from a randomly initialized condition, BPTT performs iterative gradient ascent on $\phi_{\mathrm{c}}$. While conceptually straightforward, BPTT is susceptible to vanishing and exploding gradients over long horizons.

\textbf{Cross-Entropy Method (CEM).} CEM~\cite{rubinstein1999cross} is a representative gradient-free genetic algorithm. It maintains a parametric distribution over the condition space and iteratively fits it to the top-performing samples. At each generation, a population of candidate conditions is sampled, evaluated through the surrogate model, and the elite fraction is used to update the distribution parameters.

All baselines use the same trained surrogate models. The key differentiator is how each method navigates the high-dimensional condition space: RS explores blindly, CEM uses population-based optimization, BPTT uses first-order gradients, and ZODIAC leverages the diffusion prior with compositional energy-guided denoising.

\subsection{Main Experimental Results and Analysis}
\label{sec:main_results}

\begin{table}[tb]
\centering
\begin{tabular}{lcccc}
\toprule
Method & TPR@10 & TPR@20 & TPR@50 & Spearman $\rho$ \\
\midrule
ZODIAC-d & \textbf{100.0\textsubscript{$\pm$0.0}} & \textbf{91.0\textsubscript{$\pm$5.5}} & 52.4\textsubscript{$\pm$6.3} & \textbf{0.53\textsubscript{$\pm$0.03}} \\
CEM-d & 95.0\textsubscript{$\pm$7.1} & 82.0\textsubscript{$\pm$8.4} & 41.6\textsubscript{$\pm$5.2} & 0.34\textsubscript{$\pm$0.06} \\
BPTT-d & \textbf{100.0\textsubscript{$\pm$0.0}} & 85.5\textsubscript{$\pm$9.2} & \textbf{53.2\textsubscript{$\pm$5.9}} & 0.25\textsubscript{$\pm$0.04} \\
Random-d & 12.0\textsubscript{$\pm$8.4} & 11.5\textsubscript{$\pm$6.3} & 15.8\textsubscript{$\pm$4.9} & 0.03\textsubscript{$\pm$0.07} \\
\hline
ZODIAC-ind & \textbf{97.0\textsubscript{$\pm$4.8}} & \textbf{82.5\textsubscript{$\pm$7.2}} & \textbf{44.6\textsubscript{$\pm$7.1}} & \textbf{0.50\textsubscript{$\pm$0.03}} \\
CEM-ind & 88.0\textsubscript{$\pm$11.4} & 74.0\textsubscript{$\pm$9.9} & 33.0\textsubscript{$\pm$4.6} & 0.31\textsubscript{$\pm$0.07} \\
BPTT-ind & 91.0\textsubscript{$\pm$11.0} & 77.5\textsubscript{$\pm$11.8} & 35.8\textsubscript{$\pm$4.8} & 0.21\textsubscript{$\pm$0.05} \\
Random-ind & 15.0\textsubscript{$\pm$10.8} & 12.0\textsubscript{$\pm$7.9} & 13.2\textsubscript{$\pm$5.7} & 0.04\textsubscript{$\pm$0.08} \\
\hline
ZODIAC-imp & \textbf{93.0\textsubscript{$\pm$8.2}} & \textbf{78.0\textsubscript{$\pm$9.1}} & \textbf{38.2\textsubscript{$\pm$8.5}} & \textbf{0.46\textsubscript{$\pm$0.04}} \\
CEM-imp & 82.0\textsubscript{$\pm$13.0} & 68.5\textsubscript{$\pm$11.7} & 28.4\textsubscript{$\pm$6.1} & 0.28\textsubscript{$\pm$0.08} \\
BPTT-imp & 86.0\textsubscript{$\pm$12.4} & 71.0\textsubscript{$\pm$13.2} & 30.6\textsubscript{$\pm$7.0} & 0.19\textsubscript{$\pm$0.06} \\
Random-imp & 8.0\textsubscript{$\pm$7.6} & 7.5\textsubscript{$\pm$6.1} & 9.8\textsubscript{$\pm$5.2} & 0.01\textsubscript{$\pm$0.06} \\
\hline
ZODIAC-ns3 & \textbf{90.0\textsubscript{$\pm$9.4}} & \textbf{75.0\textsubscript{$\pm$10.3}} & \textbf{36.8\textsubscript{$\pm$9.2}} & \textbf{0.43\textsubscript{$\pm$0.05}} \\
CEM-ns3 & 78.0\textsubscript{$\pm$14.3} & 65.0\textsubscript{$\pm$12.5} & 26.2\textsubscript{$\pm$6.8} & 0.26\textsubscript{$\pm$0.07} \\
BPTT-ns3 & 83.0\textsubscript{$\pm$12.8} & 68.5\textsubscript{$\pm$14.0} & 29.0\textsubscript{$\pm$7.4} & 0.17\textsubscript{$\pm$0.06} \\
Random-ns3 & 6.0\textsubscript{$\pm$6.5} & 5.5\textsubscript{$\pm$4.8} & 8.4\textsubscript{$\pm$4.3} & $-$0.01\textsubscript{$\pm$0.08} \\
\bottomrule
\end{tabular}
\caption{TPR@Top-K results for each conflict scenario. Bold values indicate the best performance.}
\label{tab:main}
\vspace{-0.75cm}
\end{table}

We conducted the experiment over 5 random seeds and present the TPR as the main result in Table~\ref{tab:main}. For each scenario, there are 1000 initial states evaluated. Several key observations emerge from these results. First, all optimization-based methods substantially outperform RS, confirming that active search over the condition space is essential for reliably discovering conflict-inducing scenarios. It is noteworthy that the conflict-related hyperparameters are specifically tuned to maintain about 10\% TPR@10 for random search, reflecting the low base rate of conflicts under natural conditions.

While CEM and BPTT achieve comparable TPR scores to ZODIAC at Top-10 and Top-20, the critical differentiator is the Spearman rank correlation. ZODIAC achieves an average $\rho = 0.48$, which is substantially higher than CEM and BPTT. This indicates that the energy produced by ZODIAC's compositional guidance not only identifies conflict scenarios but also reasons about these conflicts, thereby ensuring that the generated scenarios align with the physical environment; in contrast, while CEM and BPTT are indeed capable of discovering conflict conditions, the correlation between their estimated scenarios and the ground truth remains notably weak, mainly due to the cumulative error.

\begin{figure}[htbp]
    \centering
    \includegraphics[width=\linewidth]{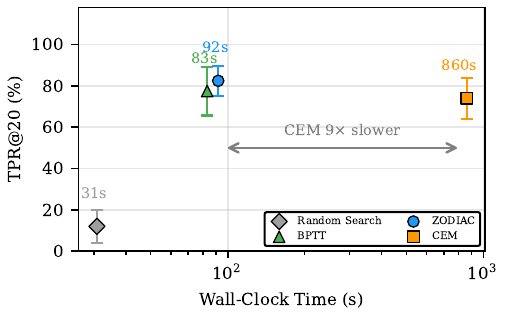}
    \caption{The average efficiency over all environments.}
    \label{fig:pareto}
\end{figure}

\begin{figure}[htbp]
    \centering
    \includegraphics[width=0.9\linewidth]{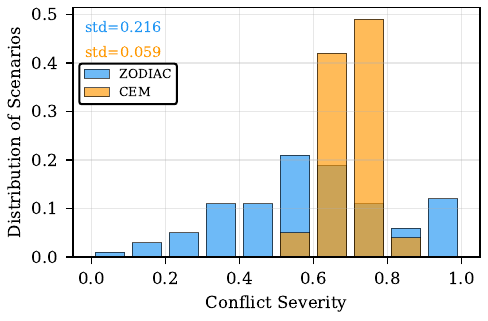}
    \caption{Diversity comparison for indirect conflict case.}
    \label{fig:diversity}
\end{figure}

Furthermore, we analyze and report the comparison of methods in efficiency and diversity. As demonstrated in Figure~\ref{fig:pareto} and Figure~\ref{fig:diversity}, ZODIAC shows competitive efficiency compared to BPTT, while both run significantly faster than CEM. As for diversity, we checked the distribution of the severity of all identified conflicts, defined as the proportion of conflict-manifesting time in the indirect conflict case. The results imply that CEM tends to fall into several modes, while ZODIAC can generate more diversified scenarios.

% \subsubsection{Role of the Diffusion Prior}

% The diffusion model serves a dual purpose in ZODIAC. First, it provides a \emph{physical prior} that constrains the generated trajectories to lie on the manifold of plausible system evolutions, preventing the guided search from drifting into physically impossible regions of the state space. Second, it enables \emph{parallel gradient-based optimization} over entire trajectories simultaneously, as opposed to the serial gradient propagation in BPTT. The superiority of ZODIAC's Spearman correlation over BPTT ($0.491$ vs.\ $0.097$ in the indirect scenario) demonstrates that the diffusion prior's regularization effect leads to better-calibrated conflict severity estimates, even when both methods find a comparable number of conflicts.

% \subsubsection{Scalability Considerations}

% The computational cost of ZODIAC is dominated by the reverse diffusion process with guided denoising steps. In our experiments, ZODIAC requires approximately 69 seconds to generate 200 candidate scenarios for the indirect conflict case, which is comparable to BPTT (67.5s) and substantially faster than exhaustive methods. CEM achieves faster wall-clock time (4.4s) due to its simpler per-iteration computation but sacrifices ranking quality ($\rho = 0.208$). The framework's cost scales linearly with the number of denoising steps and the trajectory length, making it practical for operational conflict assessment workflows where a comprehensive scan of potential conflict conditions is needed before deploying new xApp combinations.

\subsection{Ablation Studies}

To validate the contribution of each component in the compositional guidance of ZODIAC, we conduct ablation studies by systematically removing individual guidance terms and measuring the resulting degradation in performance. We consider the following variants:

\begin{itemize}
    \item \textbf{w/o $U_{\mathrm{dyn}}$}: The dynamics uncertainty penalty $\mathcal{E}_{\mathrm{dyn}}$ is removed, and the guided search is now free to exploit regions where the dynamics model produces unreliable predictions.

    \item \textbf{w/o $U_{\pi}$}: The policy uncertainty penalty $\mathcal{E}_{\pi}$ is removed, without which the inferred actions in highly out-of-distribution states become unreliable.

    \item \textbf{w/o $U_{\mathrm{dyn}} + U_{\pi}$}: Remove both penalties.

    \item \textbf{w/o Guidance}: The entire guidance mechanism is disabled, and the diffusion model generates trajectories purely from its learned unconditional prior $p_\theta(\tau)$. However, as the ranking mechanism through simulated conflict metrics is still functioning, it's still better than random sampling.  
    
\end{itemize}

\begin{figure}
    \centering
    \includegraphics[width=0.85\linewidth]{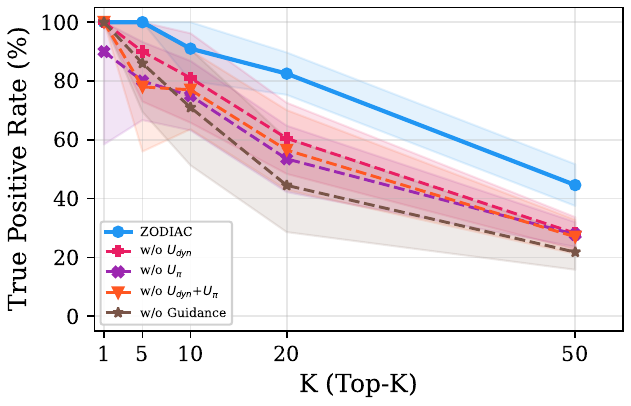}
    \caption{Visualization of the conflict cases for three different types in Mobile-Env environment.}
    \label{fig:ablation}
    \vspace{-0.1cm}
\end{figure}

\begin{figure}
    \centering
    \includegraphics[width=0.85\linewidth]{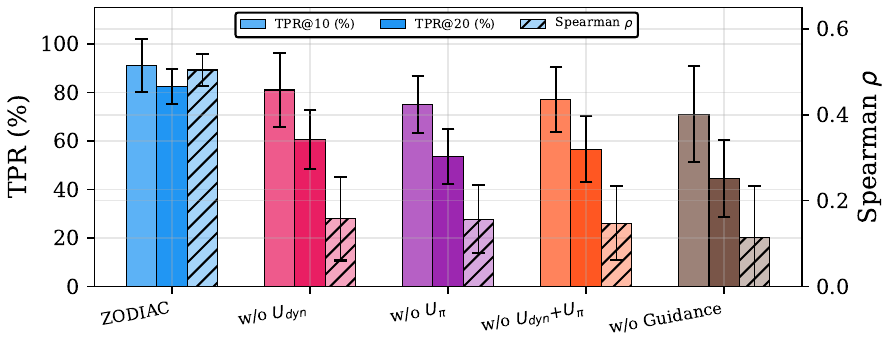}
    \caption{Visualization of the conflict cases for three different types in Mobile-Env environment.}
    \label{fig:ablation_b}
    \vspace{-0.2cm}
\end{figure}

The results shown in Figure~\ref{fig:ablation} and Figure~\ref{fig:ablation_b} reveal a clear hierarchy of component importance. The most significant degradation occurs when the entire guidance mechanism is removed (without Guidance), with TPR@10 dropping to about 70\% and TPR@20 dropping to about 50\%. This confirms that the compositional energy-guided search is the primary driver of ZODIAC's conflict discovery capability. Removing both uncertainty penalties simultaneously (w/o $U_{\mathrm{dyn}} + U_{\pi}$) causes a notable decline, indicating that the epistemic uncertainty components collectively play a substantial role in filtering out spurious conflicts. Individually, removing $U_{\mathrm{dyn}}$ has a slightly larger impact than removing $U_{\pi}$, suggesting that dynamics model reliability is more critical than policy confidence in this scenario. This aligns with the intuition that the indirect conflict mechanism relies heavily on accurate state transition predictions.

\label{sec:ablation}

\section{Conclusion}
In this paper, we formalized the zero-shot multi-policy conflict reasoning problem for O-RAN and proposed ZODIAC, a framework that discovers conflict-inducing conditions from only marginal offline datasets without any joint execution data. ZODIAC combines uncertainty-aware surrogate model training, trajectory-level diffusion priors, and compositional guidance to generate physically plausible conflict scenarios given a specific conflict criterion. Experiments on both Mobile-Env and NS-O-RAN-FlexRIC demonstrate that ZODIAC consistently outperforms the RS, CEM, and BPTT baselines, achieving the highest TPR@K and a notably superior Spearman rank correlation (average $\rho$ = 0.48 vs $\leq 0.34$), indicating that ZODIAC not only identifies conflicts but also correctly ranks their severity. Ablation studies confirm the necessity of each guidance component, with the uncertainty penalties proving essential for filtering out spurious conflicts caused by model errors. 

By enabling conflict reasoning prior to joint deployment, ZODIAC provides a practical tool for safety auditing in multi-vendor O-RAN ecosystems and complements existing detection-mitigation pipelines. Future work includes extending the framework to support large-scale multi-xApp concurrent environments, incorporating temporal credit assignment to pinpoint the onset of conflicts within long trajectories, and validating it on operational network traces.

% \begin{acks}
% \end{acks}

\bibliographystyle{ACM-Reference-Format}
\bibliography{bibliography}

\appendix

\appendix
\section{Assumptions and Proofs of Section~\ref{sec:theory}}
\label{app:proofs}

\subsection*{A.1\quad Canonical Decomposition and Boundary Condition}

In the decomposition \eqref{eq:decomp} we set
$f_0(s,e):=F^*(s,e,\pi^0_A(s),\pi^0_B(s))$,
$f_A(s,e,a^A):=F^*(s,e,a^A,\pi^0_B(s))-f_0(s,e)$, $f_B$ analogously,
and the interaction residual $f_{AB}$ is fixed by
\eqref{eq:decomp} as an identity. By construction, $f_{AB}$ vanishes
whenever \emph{either} action equals its baseline:
\begin{equation}
f_{AB}(s,e,\pi^0_A(s),a^B)=f_{AB}(s,e,a^A,\pi^0_B(s))=0.
\label{eq:fAB-boundary}
\end{equation}

\subsection*{A.2\quad Assumptions}

\begin{description}[leftmargin=1.4em,itemsep=2pt,topsep=2pt]
  \item[\textbf{(A1) Bounded coupling.}] There exists $L_{AB}\!<\!\infty$
        such that \\ $\|f_{AB}(s,e,a^A,a^B)\|_2\le L_{AB}$ for all
        $(s,e,a^A,a^B)$.
  \item[\textbf{(A2) Lipschitz dynamics and policies.}] The true joint
        dynamics $F^*$ is $L_F$-Lipschitz in $s$ and $L_a$-Lipschitz
        in $(a^A,a^B)$; the target policies $\pi_A,\pi_B$ are
        $L_\pi$-Lipschitz in $s$.
  \item[\textbf{(A3) Lipschitz conflict.}] The single-step conflict
        metric $\phi_c$ is $L_c$-Lipschitz in $(s,a^A,a^B)$ and
        bounded by $|\phi_c|\le\phi_{\max}$.
  \item[\textbf{(A4) Calibrated ensembles.}] With probability at
        least $1-\delta$, the marginal-mean prediction errors of the
        dynamics and policy ensembles at any in-distribution query
        are bounded by \\
        $\|\hat F(s,e,a^A,a^B)-(f_0+f_A+f_B)(s,e,a^A,a^B)\|_2
        \le \beta_T\sqrt{U_{dyn}}$ \\ and
        $\|\hat\pi_k(s)-\pi_k(s)\|_2\le \beta_\pi\sqrt{U_\pi}$
        for $k\in\{A,B\}$.
\end{description}
(A1) is the only non-statistical assumption.(A2)--(A3) are
standard regularity conditions. (A4) is a mild concentration
property routinely adopted for deep ensembles. The aleatoric noise
$\xi_t$ in \eqref{eq:decomp} has $\mathbb{E}\|\xi_t\|_2\le\sigma_\xi$.
For the purpose of the theoretical analysis, we assume the policies output deterministic expected action vectors, such that $a_t^A=\pi_A(s_t)$, allowing the application of standard $L_2$ Lipschitz bounds.
%\shu{<- Here, we actually mix the action vector $a$ with probability distribution $\pi$. So I add the clarification here.}

\subsection*{A.3\quad Proof of Lemma~\ref{lem:nonid}}
A trajectory in $\mathcal{D}_A$ is generated under
$a^A_t\sim\pi_A(\cdot|s_t)$ and $a^B_t\sim\pi^0_B(\cdot|s_t)$.
Substituting into Eq.~\eqref{eq:decomp},
\begin{align*}
s_{t+1} &= f_0(s_t,e_t) + f_A(s_t,e_t,a^A_t) + f_B(s_t,e_t,\pi^0_B(s_t)) \\
        &\quad + f_{AB}(s_t,e_t,a^A_t,\pi^0_B(s_t)) + \xi_t \\
        &= f_0(s_t,e_t) + f_A(s_t,e_t,a^A_t) + 0 + 0 + \xi_t,
\end{align*}
where the two zeros follow respectively from the definition of $f_B$
at the baseline ($f_B(s,e,\pi^0_B(s))=F^*(s,e,\pi^0_A(s),\pi^0_B(s))-f_0=0$)
and from \eqref{eq:fAB-boundary}. The expression depends on neither
$f_{AB}$ nor $\tilde f_{AB}$, so replacing one by the other induces
the identical conditional law. Iterating over $t=0,\dots,T-1$ shows
that the trajectory distribution on $\mathcal{D}_A$ is invariant
under $f_{AB}\!\to\!\tilde f_{AB}$. The same argument applies to
$\mathcal{D}_B$, establishing non-identifiability. \qed

\subsection*{A.4\quad Proof of Theorem~\ref{thm:traj}}

Let $\Delta_t:=\hat s_t-s_t$ with $\Delta_0=0$. The composed
surrogate is $\hat F=f_0+f_A+f_B$ (it implicitly sets
$f_{AB}\!\equiv\!0$). Using \eqref{eq:decomp},
\begin{align}
\Delta_{t+1}
&= \hat F(\hat s_t,e_t,\hat a^A_t,\hat a^B_t)
   - F^*(s_t,e_t,a^A_t,a^B_t) - \xi_t \notag\\
&= \underbrace{[\hat F(\hat s_t,e_t,\hat a^A_t,\hat a^B_t)
              -\hat F(s_t,e_t,a^A_t,a^B_t)]}_{(I)\;\text{Lipschitz drift}} \notag\\
&\quad + \underbrace{[\hat F(s_t,e_t,a^A_t,a^B_t)
              - F^*(s_t,e_t,a^A_t,a^B_t)]}_{(II)\;\text{model bias}} - \xi_t.
\label{eq:delta-decomp}
\end{align}

\paragraph{Term (II): model bias.}
Add and subtract the true marginal-sum $f_0+f_A+f_B$:
\[
(II) = \underbrace{[\hat F-(f_0{+}f_A{+}f_B)]}_{=:\,e_T(t)}
     + \underbrace{[(f_0{+}f_A{+}f_B) - F^*]}_{=\,-\,f_{AB}}.
\]
By (A4), $\|e_T(t)\|_2\le \beta_T\sqrt{U_{dyn}^{(t)}}$ with
probability $1-\delta$. By (A1), $\|f_{AB}\|_2\le L_{AB}$. Hence
\begin{equation}
\|(II)\|_2 \le \beta_T\sqrt{U_{dyn}^{(t)}} + L_{AB}. \label{eq:II}
\end{equation}

\paragraph{Term (I): Lipschitz drift.}
By (A2),
\[
\|(I)\|_2 \le L_F\|\Delta_t\|_2 + L_a(\|\hat a^A_t-a^A_t\|_2 + \|\hat a^B_t-a^B_t\|_2).
\]
For each action discrepancy, add and subtract $\pi_A(\hat s_t)$:
\begin{align*}
\|\hat a^A_t-a^A_t\|_2
&\le \|\hat a^A_t-\pi_A(\hat s_t)\|_2 + \|\pi_A(\hat s_t)-\pi_A(s_t)\|_2 \\
&\le \beta_\pi\sqrt{U_\pi^{(t)}} + L_\pi\|\Delta_t\|_2,
\end{align*}
where the first term is the policy-ensemble error (A4) and the
second is the policy Lipschitz property (A2). The same bound holds
for $\hat a^B_t$. Substituting,
\begin{equation}
\|(I)\|_2 \le (L_F+2L_aL_\pi)\|\Delta_t\|_2
            + 2L_a\beta_\pi\sqrt{U_\pi^{(t)}}.
\label{eq:I}
\end{equation}

\paragraph{Recursion.}
Combining \eqref{eq:I}, \eqref{eq:II}, and
$\mathbb{E}\|\xi_t\|_2\le\sigma_\xi$ in \eqref{eq:delta-decomp}, with
$L':=L_F+2L_aL_\pi$ and $r_t(u)$ as in \eqref{eq:r},
\[
\mathbb{E}\|\Delta_{t+1}\|_2 \le L'\,\mathbb{E}\|\Delta_t\|_2 + r_t(u).
\]
Unrolling with $\Delta_0=0$:
\[
\mathbb{E}\|\Delta_t\|_2
\le \sum_{k=0}^{t-1}(L')^{t-1-k}\,r_k(u).
\]
A union bound over $t=1,\dots,T$ steps and the two ensemble error
sources (dynamics and policy) yields the stated probability
$1-2T\delta$. \qed

\subsection*{A.5\quad Proof of Theorem~\ref{thm:lcb}}

By (A3) and the triangle inequality,
\begin{align*}
|\hat J(u)-J^*(u)|
&\le \sum_{t=0}^{T-1}|\phi_c(\hat s_t,\hat a^A_t,\hat a^B_t)
                     -\phi_c(s_t,a^A_t,a^B_t)| \\
&\le L_c\sum_{t=0}^{T-1}\bigl(\|\Delta_t\|_2
   +\|\hat a^A_t-a^A_t\|_2+\|\hat a^B_t-a^B_t\|_2\bigr).
\end{align*}
Reusing the action-discrepancy bound from the proof of
Theorem~\ref{thm:traj},
\[
\|\hat a^A_t-a^A_t\|_2+\|\hat a^B_t-a^B_t\|_2
\le 2\beta_\pi\sqrt{U_\pi^{(t)}} + 2L_\pi\|\Delta_t\|_2,
\]
hence
\[
|\hat J-J^*|
\le L_c\!\sum_{t=0}^{T-1}\!\bigl[(1{+}2L_\pi)\|\Delta_t\|_2
   + 2\beta_\pi\sqrt{U_\pi^{(t)}}\bigr].
\]
Substituting the bound on $\|\Delta_t\|_2$ from
Theorem~\ref{thm:traj} produces the closed form
\[
|\hat J-J^*|
\le L_c\!\sum_{t=0}^{T-1}\!\Bigl[
 2\beta_\pi\sqrt{U_\pi^{(t)}}
+(1{+}2L_\pi)\!\!\sum_{k=0}^{t-1}\!(L')^{t-1-k} r_k(u)\Bigr]
=R(u).
\]
The lower bound $J^*(u)\ge \hat J(u)-R(u)$ follows immediately. The
$1-2T\delta$ probability is inherited from Theorem~\ref{thm:traj}.
\qed

\subsection*{A.6\quad Discussion of Corollary~\ref{cor:guidance}}

The composite guidance vector $d_{\mathrm{guide}}$ in Eq.~(16)
ascends a weighted combination of the three energies. Identifying
this ascent with the gradient of an implicit objective
\[
\mathcal{J}(u)\;:=\;\mathcal{E}_{\mathrm{target}}(u)
- \tfrac{1-\alpha}{\alpha}\bigl[-\mathcal{E}_{\mathrm{physics}}(u)
- \mathcal{E}_{\mathrm{epistemic}}(u)\bigr],
\]
we observe that:
\begin{enumerate}[itemsep=1pt,topsep=2pt]
\item $\mathcal{E}_{\mathrm{target}}$ contains
$\sum_t\phi_c(\hat s_t,\hat a^A_t,\hat a^B_t)=\hat J(u)$ together
with the policy log-likelihoods that constrain
$\hat a^A_t,\hat a^B_t$ to the marginal action manifold;
\item $\mathcal{E}_{\mathrm{epistemic}}$ accumulates
$U_{dyn}^{(t)}+U_\pi^{(t)}$, the only \emph{variable} component of
$R(u)$ (the Lipschitz constants and $L_{AB}$ are fixed by the
problem);
\item $\mathcal{E}_{\mathrm{physics}}$ enforces self-consistency of
the generated trajectory with the surrogate dynamics, which is a
necessary condition for (A4) to remain informative along the
rollout.
\end{enumerate}
The mapping is first-order: $R(u)$ uses $\sqrt{U}$ while ZODIAC
penalizes $U$ directly, but both are monotone and both vanish on the
same in-distribution manifold, so the maximizer of
$\hat J(u)-\lambda R(u)$ and that of $\mathcal{J}(u)$ coincide up
to the choice of $\lambda$ (controlled by $\alpha$ in Eq.~16).
Substituting any $u^\star$ returned by the diffusion sampler into
\eqref{eq:lcb} yields the certificate \eqref{eq:certificate}. \qed

\end{document}